\documentclass[sigconf, nonacm]{acmart}
\usepackage[show]{chato-notes}
\usepackage{amsfonts}
\usepackage{acmart-taps}
\usepackage{fontawesome}
\usepackage{xcolor}
\usepackage{framed}
\usepackage{graphicx}
\usepackage{float}
\usepackage{makecell}

\AtBeginDocument{%
  }

\usepackage{hyperref}
\usepackage{float}
\usepackage{svg}

\graphicspath{ {./images/} }
\begin{document}

\title[Gryphon-v2]{Gryphon-v2: One Model in Place of a Cascade --- Generate-and-Rank Recommender with Rollout Distillation}

\author{Anna Lipkina}
\authornote{These authors contributed equally to this research and are listed alphabetically.}
\affiliation{%
  \institution{Yandex}
  \city{Moscow}
  \country{Russia}
}
\email{zalipan@yandex-team.ru}

\author{Daria Tikhonovich}
\authornotemark[1]
\affiliation{%
  \institution{Yandex}
  \city{Moscow}
  \country{Russia}
}
\email{daria.m.tikhonovich@gmail.com}

\author{Viktor Yanush}
\authornotemark[1]
\affiliation{%
  \institution{Yandex}
  \city{Moscow}
  \country{Russia}
}
\email{yanushv@yandex-team.ru}

\author{Mariia Ulianova}
\orcid{0009-0004-7571-2325}
\affiliation{%
  \institution{Yandex}
  \city{Moscow}
  \country{Russia}
}
\email{devchatay@gmail.com}

\author{Oleg Sorokin}
\orcid{0009-0009-1400-2449}
\affiliation{%
  \institution{Yandex}
  \city{Moscow}
  \country{Russia}
}
\email{olegsorokin.ai@gmail.com}

\author{Vladislav Dodonov}
\orcid{0009-0009-6112-8415}
\affiliation{%
  \institution{Yandex}
  \city{Moscow}
  \country{Russia}
}
\email{golafotino@gmail.com}

\author{Ilya Murzin}
\affiliation{%
  \institution{Yandex}
  \city{Moscow}
  \country{Russia}
}
\email{ilyamurzin@yandex-team.ru}

\author{Denis Burshtein}
\affiliation{%
  \institution{Yandex}
  \city{Moscow}
  \country{Russia}
}
\email{deburo@yandex-team.ru}

\author{Nikolay Savushkin}
\affiliation{%
  \institution{Yandex}
  \city{Moscow}
  \country{Russia}
}
\email{penguin-diver@yandex-team.ru}

\renewcommand{\shortauthors}{Lipkina et al.}

\begin{abstract}

Industrial recommender systems are commonly deployed as multi-stage cascades with separate candidate generators, pre-rankers, and final rankers. Although effective, these cascades require repeated user-history processing, complex feature pipelines, and multiple serving stages. Semantic-ID-based generative retrieval offers a path toward simpler end-to-end systems, but next-item prediction alone does not capture the fine-grained preferences encoded by production ranking objectives.

We present Gryphon-v2, a unified generate-and-rank architecture for end-to-end recommendation. The model encodes a user history once, generates Semantic-ID candidates with an autoregressive decoder, resolves them to catalogue items, and ranks them with an item-level Ranking Module that reuses the shared encoder states. To transfer fine-grained production ranking preferences without adding an expensive second model to the serving path, we distill a high-capacity, training-only Teacher Ranker into the Ranking Module. Gryphon-v2 is trained with \emph{Rollout Distillation}: teacher scores are the only ranking supervision, and they are collected over two complementary candidate distributions. Rollouts from the current decoder expose the Ranking Module to candidates produced by the same generation mechanism used at serving time, while logged impressions cover items users were actually shown.

In an online A/B experiment on a large-scale recommendation surface at Yandex Music, a single Gryphon-v2 model replaces a production cascade comprising more than 15 candidate generators, pre-ranking, and final ranking. The deployment increases the number of active users by 1.41\% at serving latency comparable to the production cascade. These results support the practical viability of a generative retriever with a Ranking Module distilled from the Teacher Ranker as an end-to-end alternative to a production cascade.

\end{abstract}



\maketitle

\section{Introduction}

Industrial recommender systems typically use cascades of candidate generators, pre-rankers, and final rankers. This design is effective but costly to maintain, requiring repeated history processing, complex feature pipelines, and multiple serving stages. Generative recommendation with Semantic IDs (SIDs)~\cite{rajput2023} can consolidate parts of this pipeline, but next-item prediction alone does not capture the fine-grained, multi-objective preferences optimized by a production ranker.

Two lines of industrial work
address this gap. The first adds a post-training stage that aligns the generator
with service-level objectives through reinforcement learning against a learned
reward model or against user feedback
directly~\cite{onerec2025,onerecv22025,gpr2025,oxygenrec2025}. The second unifies candidate
generation and ranking inside a single architecture, so that item-level scoring
rather than retrieval scores determines the final
order, while the shared encoder provides efficient serving~\cite{li2026unipinrec,recochain2026,sun2026grank,tikhonovich2026gryphon}.

\begin{figure*}[!htbp]
    \centering
    \includegraphics[width=\textwidth]{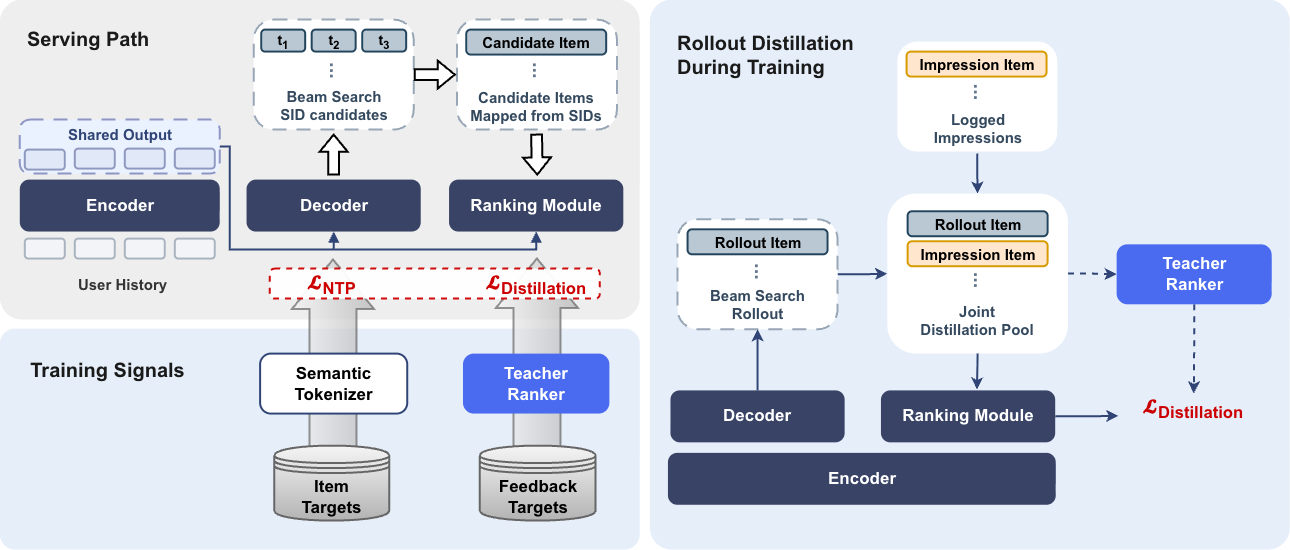}
    \caption{Gryphon-v2 framework: a unified generate-and-rank architecture that replaces the production cascade with one served model. The model is jointly trained with the next-token-prediction loss and the distillation losses. More than 90\% of distinct distillation candidates come from model beam-search rollouts.}
    \label{fig:Gryphon}
\end{figure*}

Gryphon-v2 brings together ideas from these two research directions (Figure~\ref{fig:Gryphon}). From unified generate-and-rank models, we adopt Gryphon's shared-encoder architecture~\cite{tikhonovich2026gryphon}, in which a lightweight item-level Ranking Module reorders generated candidates without re-encoding the user history. From reward-based post-training, we adopt the principle of transferring fine-grained service-level preferences from a high-capacity offline model to the served model. We implement this transfer by distilling a Teacher Ranker—validated in an online A/B experiment to outperform the production ranker, but too computationally expensive to serve at full traffic scale—into the Ranking Module. The teacher therefore plays a role analogous to a reward model, while the optimization remains supervised distillation rather than reinforcement learning.


To train the integrated Ranking Module, Gryphon-v2 uses \emph{Rollout Distillation}, a procedure inspired by on-policy distillation in LLM post-training~\cite{agarwal2024policy,li2026rethinking}. A training-only Teacher Ranker supplies preference targets for candidates drawn from two complementary sources: rollouts generated by the current SID decoder using the same constrained-decoding mechanism as at serving time, and logged impressions, which broaden coverage to items exposed in production traffic.
In an online A/B experiment on a large-scale music recommendation surface at Yandex Music, Gryphon-v2 replaces the full production cascade --- more than 15 candidate generators, pre-ranking, and final ranking --- with one served model, improving the number of active users by 1.41\% at comparable serving latency.

In summary, our main contributions are the following:
\begin{itemize}

\item  We introduce Gryphon-v2 --- a unified generate-and-rank recommender in which SID generation provides high-recall candidates and an item-level Ranking Module, distilled from a high-capacity, training-only Teacher Ranker, produces the final item ordering.

\item We develop \emph{Rollout Distillation}, a training procedure in which the teacher scores candidates generated by the current decoder --- the same decoding mechanism used at serving time --- together with logged impressions as a complementary candidate source.

\item We report an online A/B experiment in which a single served generate-and-rank model replaces a mature production cascade, while significantly improving user-engagement metrics at serving latency comparable to the cascade.

\end{itemize}

\section{Related Work}
\label{sec:background}

\subsection{End-to-End Generative Recommendation}

Multi-stage recommender systems typically optimize cascade stages with
stage-specific objectives, which makes end-to-end optimization of the final
recommendation outcome difficult. Generative recommendation addresses this
fragmentation by conditioning one model on the user's interaction history and
directly generating Semantic IDs (SIDs) of recommended
items~\citep{rajput2023}.

Several deployed generative recommenders augment next-token pre-training with
post-training against service-level objectives. OneRec applies on-policy
reinforcement learning using a learned preference reward model
\citep{onerec2025}; OneRec-V2 uses rewards derived directly from real user
feedback together with a lazy decoder-only architecture that removes the
encoder and simplifies cross-attention \citep{onerecv22025}; GPR combines
multi-token and value-aware training with hierarchical policy optimization
\citep{gpr2025}; and OxygenREC combines near-line reasoning instructions with
a unified ranking-model reward service and scenario-aware policy optimization
\citep{oxygenrec2025}. Related
systems apply reward- or preference-based post-training in local services,
e-commerce, live-streaming, and advertising
\citep{oneloc2025,onemall2026,onelive2026,qiu2025,zheng2025}. Across this
family, the shared pattern is an explicit post-training stage for the
generative policy; the supervision itself varies among learned reward or
ranking models, direct user-feedback rewards, and task-specific reward
functions.

In these formulations, post-training modifies the generator that produces the
served candidates or slate. Gryphon-v2 uses Semantic-ID generation as a
high-recall proposal mechanism, while its final item order is produced by a
Ranking Module that shares the history encoder with the generator and is
distilled from a training-only Teacher Ranker.

\subsection{Unified Architecture for Retrieval and Ranking}

A second line of work couples candidate generation and ranking objectives within a single model~\cite{sun2026unisgr,gao2025synergen}. Existing methods typically reuse history encodings across stages or condition ranking on generated candidates.

UniPinRec unifies retrieval and ranking tasks by using a shared input format, joint training, and a common backbone, while still serving retrieval through approximate nearest-neighbor search over item embeddings~\cite{li2026unipinrec}.
GRank pairs a MIPS-based generator trained under target-aware supervision with a lightweight cross-attention ranker. It avoids structured item-centric indexes, but still relies on maximum-inner-product search and maintains separate user-history encoders for candidate generation and ranking~\cite{sun2026grank}.
RecoChain merges SID generation and reranking within a single causal decoder, while relying on searched sequence modeling against the generated target and using distinct, stage-specific input formats for generation and reranking~\cite{recochain2026}.
UniSGR introduces value-aware SID generation together with a unified multi-objective ranking module, aligning generation with click, cart, purchase, and value-oriented objectives~\cite{sun2026unisgr}.

Gryphon identifies a mismatch between SID generation and item-level scoring in generative recommendation. It proposes a shared-encoder architecture with an Item-Level Scoring Module, decoupling retrieval-time item selection from potentially miscalibrated SID beam scores~\cite{tikhonovich2026gryphon}.

We build Gryphon-v2 upon the Gryphon architecture, introducing distillation training of the item-level Ranking Module and allowing a single model to replace the full production cascade from retrieval to final ranking.

\subsection{Distillation in Recommender Systems}

Knowledge distillation trains a student to reproduce the signals of a teacher
model~\cite{hinton2015distilling}. In recommendation it is a standard tool for
transferring ranking quality, compressing expensive scorers, and aligning
retrieval-stage models with objectives defined later in the
cascade~\cite{tang2018rankingdistillation,lee2019collaborative,zhu2020ensembled}. Continuous teacher scores are
especially informative for ranking, preserving the preference intensity and
ordering that hard labels discard \cite{tang2018rankingdistillation,kang2020rrd}.

Industrial recipes decouple offline model scale from online serving cost.
Rec-Distill distills a separately trained large teacher into a lightweight
served student through black-box logit matching~\cite{recdistill2026}, and
Meta's ExFM formalizes this as \emph{external distillation}: a foundation-model
teacher trained apart from the served models supplies logged scalar predictions
as supervision, so scaling the teacher adds no serving cost~\cite{exfm2025}. To
widen the scalar bottleneck, other designs transfer teacher knowledge as
features---target-aware embeddings for lightweight experts~\cite{foundationexpert2025}
or historical embedding sequences for vertical models~\cite{loopfm2026}.

Several industrial distillation approaches draw distillation candidates from logged impressions~\cite{tang2018rankingdistillation,exfm2025,recdistill2026}, which provide stable supervision on observed traffic. Training students on model-generated outputs is, in turn, well established in LLM post-training~\cite{agarwal2024policy,li2026rethinking}. Gryphon-v2 adapts this candidate-selection principle to a deployed generate-and-rank recommender by distilling the Ranking Module on candidates from the current SID decoder, while retaining logged impressions as complementary candidate source.

\section{Methodology}\label{sec:method}
Replacing a cascade with one served model imposes two requirements that the
cascade satisfies with separate components: the model must produce a
sufficiently broad candidate set and order those candidates effectively, while processing the user history once within the serving latency budget.

Gryphon-v2 inherits Gryphon's shared-encoder architecture: SID generation
provides high-recall proposals, and an item-level Ranking Module reuses the same history
states to determine the final item order. The key difference is how the Ranking Module is supervised and deployed
(Figure~\ref{fig:Gryphon}). In Gryphon-v2, next-token prediction (NTP) trains the generative path, while the training-only Teacher Ranker provides distillation targets for the Ranking Module on decoder rollouts and logged impressions.

\subsection{Architecture Overview}

Gryphon-v2 uses Gryphon's shared-encoder
architecture~\cite{tikhonovich2026gryphon}---a history encoder,
autoregressive SID decoder, and item-level Ranking
Module.

\paragraph{Encoder.} Let \(u\) index a request and its associated user context,
and let \(H_u=(h_{u,1},\ldots,h_{u,n_u})\) denote the chronological sequence of
events available before the request-time cutoff. A
bidirectional Transformer maps this sequence to contextual states
\[
  E_u=\operatorname{Encode}_{\mathrm{hist}}(H_u).
\]
The history is encoded once. Both the decoder and the Ranking Module reuse the
same request-specific activations \(E_u\), so neither path separately
re-encodes \(H_u\).

\paragraph{Decoder.} The autoregressive decoder cross-attends to \(E_u\) and emits the SID of
the next item. At inference, catalogue-trie-constrained beam
search produces valid candidate SIDs, which are resolved to concrete
catalogue items (Section~\ref{sec:generation}). The decoder acts as a proposal
mechanism: its sequence likelihood determines beam membership, but not the
final order of resolved items.

\paragraph{Ranking Module.} For each resolved candidate, the Ranking Module uses an item representation as
a query over \(E_u\). It predicts task-specific item-level scores whose fixed
combination determines the final ordering (Section~\ref{sec:ranking}).

\subsection{Request-wise Training Instances}
\label{sec:dataset-design}

Each training instance corresponds to a served request indexed by \(u\). Let
\(\mathcal{M}_u\) be the set of catalogue items impressed in that request,
together with their associated feedback. The subset
\(\mathcal{P}_u\subseteq\mathcal{M}_u\) contains the positive impressed items
used as NTP targets. Let \(\mathcal{T}\) denote the fixed set of ranking tasks.

During joint training, the current SID decoder generates and resolves
the rollout candidate set \(\mathcal{G}_u\). The Teacher Ranker and Ranking
Module score the same candidates from \(\mathcal{M}_u\) and
\(\mathcal{G}_u\), producing a teacher target \(r^{T,t}_{u,i}\) for each
candidate \(i\) and task \(t\in\mathcal{T}\), where the superscript \(T\)
denotes the teacher. Logged and generated distillation candidates provide complementary sets for teacher scoring. Grouping these signals by request allows the student
to encode \(H_u\) once and reuse \(E_u\) for all generation and distillation targets.

\subsection{Semantic-ID Generation and Resolution}
\label{sec:generation}

\paragraph{Next-token-prediction training.}

Each catalogue item \(i\) is assigned a Semantic ID
\(\Phi(i)=(s_1,\ldots,s_L)\), a tuple of discrete codes drawn from \(L\)
hierarchical codebooks and arranged coarse-to-fine.

Conditioned on \(E_u\), the decoder factorizes an SID autoregressively and
predicts one code at each position. For every positive target
\(i^+\in\mathcal{P}_u\) with
\(\Phi(i^+)=(s_1^+,\ldots,s_L^+)\), its teacher-forced NTP contribution is
\[
\ell_{\mathrm{NTP}}(u,i^+)
=
-\sum_{\ell=1}^{L}
\log p_\theta(s_\ell^+ \mid s_{<\ell}^+, E_u).
\]
Here \(\theta\) denotes the model parameters on the NTP path. For requests with multiple positive targets, we sum their teacher-forced
contributions:
\[
\mathcal{L}_{\mathrm{NTP}}(u)
=
\sum_{i^+\in\mathcal{P}_u}
\ell_{\mathrm{NTP}}(u,i^+).
\]

\paragraph{Constrained generation.}
During inference, catalogue-trie-constrained beam
search ensures that every completed SID resolves to at least one catalogue item. Using beam search with beam size \(K\), the decoder returns an ordered SID
beam \(\mathcal{B}^{(K)}_u\).

\paragraph{Collision resolution and candidate budgeting.}
Several items may correspond to the same SID. Let
\(\mathcal{C}_\sigma=\{\,i:\Phi(i)=\sigma\,\}\) denote the collision class of
a generated SID \(\sigma\). Expanding every SID in the generated beam gives
the unbudgeted candidate set
\[
\widetilde{\mathcal{I}}_u
=
\bigcup_{\sigma\in\mathcal{B}^{(K)}_u}
\mathcal{C}_\sigma.
\]

We construct the semantic index to keep collision classes small. In deployment,
the resolved candidate pool is capped at the item budget
\(B_{\mathrm{item}}\). If
\(\lvert\widetilde{\mathcal{I}}_u\rvert>B_{\mathrm{item}}\), SID hypotheses
are removed from \(\mathcal{B}^{(K)}_u\) in ascending beam-score order,
together with their collision classes, until the item budget is satisfied.
Let \(\mathcal{B}^{(K)}_{u,\mathrm{keep}}\) denote the retained SID beam. The
final candidate pool is
\[
\mathcal{I}_u
=
\bigcup_{\sigma\in\mathcal{B}^{(K)}_{u,\mathrm{keep}}}
\mathcal{C}_\sigma,
\qquad
\lvert\mathcal{I}_u\rvert\le B_{\mathrm{item}}.
\]

All items in \(\mathcal{C}_\sigma\) inherit the same beam score, so this score
cannot distinguish items within a collision class. More generally,
autoregressive SID likelihood may be poorly calibrated for item-level
relevance estimation~\cite{promise2026,tikhonovich2026gryphon}. Beam scores
are therefore used only for proposal generation and rare capacity-control
truncation. The Ranking Module determines the final ordering of
\(\mathcal{I}_u\) (Section~\ref{sec:ranking}).

\subsection{Ranking Module}
\label{sec:ranking}

The item-level Ranking Module scores resolved candidates in
\(\mathcal{I}_u\) using the shared history states \(E_u\), without re-encoding
the user history. Its candidate encoder can incorporate additional item-level
features through the candidate representation.

For candidate \(i\), let \(\mathbf{x}_i\) denote its item-level features. The
candidate encoder maps these features to a candidate representation:
\[
\mathbf e_i
=
\operatorname{Encode}_{\mathrm{item}}(\mathbf{x}_i).
\]

In the reported deployment, \(\mathbf{x}_i\) contains the item identifier and its SID
\(\Phi(i)=(s_1,\ldots,s_L)\). Both feature sources are represented using
compositional multi-hash embeddings retrieved from the unified embedding table
shared with the history encoder~\cite{shi2020compositional,coleman2023unified}.
For the SID, we construct prefix \(n\)-gram features from its code sequence
following~\cite{zheng2025enhancing}.

The candidate representation attends to the shared encoded history through one or more cross-attention blocks:
\[
  \tilde{\mathbf e}_{u,i}
  =
  \operatorname{CrossAttn}(\mathbf e_i, E_u).
\]
A task-specific output head maps \(\tilde{\mathbf e}_{u,i}\) to one score for
each ranking task \(t\in\mathcal{T}\), matching the output structure of the
Teacher Ranker (Section~\ref{sec:teacher}):
\[
\hat r^{t}_{u,i}
=
\operatorname{Head}_{t}(\tilde{\mathbf e}_{u,i}),
\qquad t\in\mathcal{T}.
\]

For the final ordering, the task scores are combined using fixed weights:
\[
\hat R(u,i)
=
\sum_{t\in\mathcal{T}} w_t\,\hat r^{t}_{u,i},
\]
where \(w_t\) is the weight assigned to task \(t\). Items in
\(\mathcal{I}_u\) are ordered by decreasing \(\hat R(u,i)\), with no
additional reranking during serving.

\subsection{Teacher Ranker}
\label{sec:teacher}

The Teacher Ranker is a large sequential model that scores candidate items
using long user interaction histories, without relying on hand-crafted tabular
features. It was selected through prior internal offline and online validation
against the production ranker. These evaluations predated the Gryphon-v2
experiment and are not included as controlled results in this paper; we use
them only to motivate the choice of distillation teacher. Its computational
cost makes direct deployment impractical, so it is used only during training
to supervise the Ranking Module through distillation.

Architecturally, it pairs the sequential history encoder with a cross-attention
ranker, where each candidate attends directly to the encoded
history~\cite{sun2026grank,tikhonovich2026gryphon}, and uses task-specific
heads for multi-objective learning.

Following ARGUS~\cite{khrylchenko2025scaling}, we train the teacher
autoregressively over chronological interaction sequences, obtaining
supervision from multiple positions within each sequence. This efficient
training formulation allows the teacher to learn from substantially more data
than the Gryphon-v2 student within a practical training budget. The teacher
also conditions on histories of up to \(8{,}000\) events, a context length that
would be prohibitively expensive for the served Gryphon-v2 model. The featureless design enables the teacher to score any user–item pair based solely on the preceding user history, allowing it to provide distillation targets for rollout candidates.

\subsection{Rollout Distillation}
\label{sec:distillation}
Rollout Distillation trains the Ranking Module using Teacher Ranker scores for
candidates from two sources: current-decoder rollouts and logged impressions. The rollout candidate-selection principle is inspired by on-policy distillation in LLM post-training~\cite{agarwal2024policy,li2026rethinking}, but the optimization itself remains supervised, and no gradient is propagated through beam search.


For a request \(u\), a candidate item
\(i \in \mathcal{G}_u \cup \mathcal{M}_u\), and a ranking task
\(t \in \mathcal{T}\), let \(r^{T,t}_{u,i}\) and \(\hat r^t_{u,i}\)
denote the Teacher Ranker and Ranking Module scores, respectively.

\paragraph{Rollout candidates.}

For each user history \(H_u\), constrained beam search
synchronously generates a beam of SIDs \(\mathcal{B}_u\) using the current
decoder parameters at every training step, with no checkpoint lag.
The generated SIDs are then
resolved into valid catalogue items, producing the rollout candidate set
\(\mathcal{G}_u\) for the current user request. The Ranking Module scores each
item conditioned on the shared encoder states \(E_u\), while the teacher scores
the same items with its own architecture. Distillation is applied independently
for each ranking task, with all tasks entering the objective with equal weight, where rollout candidates distillation loss is element-wise MAE:
\[
  \mathcal{L}_{\mathrm{roll\mbox{-}distill}}
  =
  \frac{1}{|\mathcal{T}|}
  \sum_{t\in\mathcal{T}}
  \frac{1}{|\mathcal{G}_u|}
  \sum_{i\in\mathcal{G}_u}
  \left|\hat r^t_{u,i}-r^{T,t}_{u,i}\right|.
\]
In our training data, rollout candidates account for more than
90\% of the distinct distillation candidates. This share describes candidate
coverage, not relative loss weight: each source loss is normalized by its own candidate count, and both distillation losses enter the joint
objective with equal coefficients.

 \paragraph{Impression candidates.}

We complement the rollout candidate set \(\mathcal{G}_u\) with the items \(\mathcal{M}_u\) displayed in the corresponding logged request. This pool broadens the set of teacher-scored candidates beyond current-decoder rollouts and anchors training to items observed in production traffic. We apply the same task-wise MAE objective:
\[
  \mathcal{L}_{\mathrm{impr\mbox{-}distill}}
  =
  \frac{1}{|\mathcal{T}|}
  \sum_{t\in\mathcal{T}}
  \frac{1}{|\mathcal{M}_u|}
  \sum_{i\in\mathcal{M}_u}
  \left|\hat r^t_{u,i}-r^{T,t}_{u,i}\right|.
\]

Rollout Distillation is the sum of the two candidate-source terms,
each normalized by its own candidate count and entering with
coefficient one:
\[
  \mathcal{L}_{\mathrm{distill}}
  =
  \mathcal{L}_{\mathrm{roll\mbox{-}distill}}
  +\mathcal{L}_{\mathrm{impr\mbox{-}distill}},
\]

\subsection{Joint Training}
\label{sec:joint}

We train the unified Gryphon-v2 architecture with a joint objective combining next-token prediction and Rollout Distillation:
\[
  \mathcal{L}
  =
  \mathcal{L}_{\mathrm{NTP}}
  +\mathcal{L}_{\mathrm{distill}}.
\]

The NTP loss updates the decoder and the shared encoder, while the
distillation updates the Ranking Module and the shared encoder. The
Teacher Ranker provides targets during training but is not included in the
Gryphon-v2 serving graph. Joint training transfers item-level ranking information
into the same encoded user representation \({E}_u\) used for SID generation.

\section{Deployment}
\label{sec:deploy}
Gryphon-v2 serves the traffic allocated to the online experiment as the sole
learned recommendation model on the treatment path. Serving starts from the
offline checkpoint described in Section~\ref{sec:experiments}, after which the
model receives online updates based on recent interaction data. The
Teacher Ranker supplies training targets and is refreshed daily, but is never
part of the serving path.

Online training instances are grouped by request using logged impressions and
feedback. The corresponding
history \(H_u\) is reconstructed at the request-time cutoff using the same
profile representation and feature construction as serving, reducing
training--serving skew.

The served model is updated on a ten-minute cadence, and each update itself takes on the order of tens of minutes. As a result, the model serving the experiment is typically trained on user events observed less than an hour earlier.

At request time, a CPU service constructs the request features and sends them
to Gryphon-v2, which is served on GPU through the NVIDIA Triton Inference Server. The model returns
the final item ordering without downstream learned pre-ranking or ranking
stages. Section~\ref{sec:efficiency} reports the resulting serving latency and
compares this path with the production cascade.

\section{Experiments}
\label{sec:experiments}

We organize the empirical study around three research questions.

\textbf{RQ1.} How does Gryphon-v2 compare with matched generative baselines in candidate-generation quality, fidelity to the Teacher Ranker, and ranking accuracy on logged impressions?

\textbf{RQ2.} Can Gryphon-v2 replace the production cascade --- candidate generation, pre-ranking, and ranking --- with a single generate-and-rank model while improving online engagement metrics?

\textbf{RQ3.} How do distillation loss, rollout beam size, and candidate-source composition affect Gryphon-v2 performance?

\subsection{Offline Experimental Setup}

\paragraph{Dataset.}
We construct chronological training and test sets from two weeks of
interaction logs of a large-scale music recommendation service, using the
request-wise instances defined in Section~\ref{sec:dataset-design}. 
This two-week window is used for all offline comparisons only; the online experiment starts from a four-week checkpoint and is updated online.
We apply temporal split, so
test requests occur strictly after the training window and covers the last day of logs.
The logs come from a purely recommendation-driven surface: an infinite personalized music feed with no search or editorial context. We retain a request only if at least one of its impressions is positive, where an impression counts as
positive if the track received a like or a long listen
and was not disliked. This guarantees that every training request carries at
least one positive target for next-token prediction. 


\paragraph{Baselines.}
We compare Gryphon-v2 to the following baselines:
\begin{itemize}
  \item \textbf{Generative retrieval}: the encoder--decoder backbone with the
  SID decoder alone; resolved candidates are ordered by
  SID beam likelihood. It is trained with the same Semantic-ID next-token-prediction objective as Gryphon-v2, but has no item-level Ranking Module.
  \item \textbf{Gryphon}~\cite{tikhonovich2026gryphon}: It uses the same shared encoder, SID decoder, and item-level Ranking Module as Gryphon-v2, but the Ranking Module is trained from next-item supervision rather than Teacher Ranker distillation.
\end{itemize}
Gryphon-v2 and both baselines use the same tokenizer, encoder--decoder backbone, history features, candidate budget, and constrained decoding procedure. Gryphon and Gryphon-v2 use the same item-level Ranking Module architecture, so their comparison isolates the effect of teacher distillation versus next-item supervision.

\paragraph{Evaluation protocol.}
We report three complementary diagnostics. First, Recall@1000 measures candidate-generation recall on resolved candidates before item-level reranking. Second, TeacherRecall@\(k\) measures fidelity to the Teacher Ranker's top-\(k\) selections: models evaluated with an active Ranking Module use its reranked order, while beam-order evaluations use SID beam likelihood. Third, weighted pair accuracy (WPA) measures ordering quality on production-logged impression pairs from the test set.

\paragraph{Metrics.}
Candidate generation quality is evaluated with Recall@1000: the fraction of held-out
targets for each user request recovered among the top-\(1{,}000\) resolved items. We measure Recall@1000 on the candidate set ordered strictly by beam search scores, before applying the item-level Ranking Module if it is present in the model. Tables abbreviate the metric as R@1000.

We use the Teacher Ranker's top-\(k\) selections as an offline reference for
evaluating distillation fidelity. This metric does not independently measure
recommendation quality; rather, it measures how closely the student reproduces the teacher on the student's serving-time candidate distribution.
For each user request, the model and the Teacher Ranker score the same candidate pool produced by decoder beam search at the production beam size. We define TeacherRecall@\(k\) as the fraction of the Teacher Ranker's top-\(k\) items recovered in the model's top-\(k\):

\[
\mathrm{TeacherRecall@}k(u)
=
\frac{1}{k}
\left|
\operatorname{Top}_k\!\left(s^{\mathrm{model}}_{u}\right)
\cap
\operatorname{Top}_k\!\left(s^{\mathrm{teacher}}_{u}\right)
\right|,
\]
averaged over the evaluated user requests:
\[
\mathrm{TeacherRecall@}k
=
\frac{1}{|\mathcal{U}|}
\sum_{u\in\mathcal{U}}
\mathrm{TeacherRecall@}k(u).
\]
A value of \(1.0\) indicates that the model selects the same top-\(k\) items
as the teacher from the candidate set, and \(0.0\) indicates disjoint
top-\(k\) sets. Tables abbreviate the metric as T-R@10 and T-R@100.

To complement this teacher-relative measure, we report \emph{weighted pair accuracy} (WPA), the label-based metric used to evaluate our production ranking models. The
test set contains temporally adjacent
recommendation impressions with different engagement grades. Eligible pairs
may occur within one recommendation request or across the boundary between two
adjacent requests. The engagement ordering is
\[
\text{like} > \text{play} > \text{skip} > \text{dislike}.
\]

Each engagement grade has a predefined target weight \(t\). We orient every
eligible pair \((i,j)\in\mathcal{P}\) such that \(t_i>t_j\), and compute

\[
\operatorname{WPA}
=
\frac{
  \sum_{(i,j)\in\mathcal{P}}
    (t_i-t_j)\,
    \mathbf{1}\!\left\{s(i)>s(j)\right\}
}{
  \sum_{(i,j)\in\mathcal{P}}(t_i-t_j)
},
\]
where \(s(\cdot)\) is the scalar ordering score: SID beam likelihood for
beam-order evaluations and the ranker's score for evaluations with an active
ranker. Thus,
a correctly ordered pair contributes in proportion to the difference between
its engagement weights, and pairs with more widely separated grades contribute
more strongly.

The Ranking Module, Teacher Ranker, and production ranker are evaluated on the
same impression pairs from the test set. Weighted pair accuracy uses logged engagement labels
as its target under the existing cascade's exposure policy. Tables abbreviate the metric as WPA.

\paragraph{Implementation details.}
Architecture hyperparameters and the optimization schedule are reported in
Appendix~\ref{sec:hparams-training}.
Offline training makes one chronological pass over the two-week window.
Training rollouts use beam size 32.
Ranking tasks and the three joint-objective losses are equally weighted.

We construct multimodal item representations by aligning text and audio
features with collaborative signals following QARM~\cite{luo2025qarm}. Audio and textual content features are
extracted with Qwen2.5-Omni and projected by a small Transformer trained with
an in-batch contrastive objective. Residual \(K\)-means quantization produces
the hierarchical SIDs. The resulting index has an Independent Code Rate of \(0.98\).

At inference, the decoder generates \(1{,}024\) valid SIDs. After collision
expansion, the pool is capped at \(B_{\mathrm{item}}=1{,}200\) by removing the
lowest-scoring SID hypotheses and their collision classes.
The online experiment uses the same configuration, except that the model is
initially trained on four weeks of data and subsequently updated online
(Section~\ref{sec:deploy}).

\subsection{Offline Results (RQ1)}

Table~\ref{tab:offline} reports the offline comparison on the test set.

\begin{table}[t]
  \centering
  \small
  \setlength{\tabcolsep}{3pt}
  \caption{Offline comparison of generative models: candidate-generation recall
(R@1000), agreement with the Teacher Ranker (T-R@\(k\)), and label-based
ordering accuracy (WPA).}
  \resizebox{\columnwidth}{!}{%
  \begin{tabular}{lcccc}
    \toprule
    \textbf{Model}
      & \textbf{R@1000}
      & \textbf{T-R@10}
      & \textbf{T-R@100}
      & \textbf{WPA} \\
    \midrule
    \multicolumn{5}{l}{\textit{Generative models}} \\
    Generative retrieval
      & \textbf{0.8643} & 0.0382 & 0.1944 & 0.5429 \\
    Gryphon
      & 0.8593 & 0.0392 & 0.1701 & 0.5528 \\
    Gryphon-v2 (beam order)
      & 0.8615 & 0.0381 & 0.1936 & 0.5478 \\
    Gryphon-v2 (ours)
      & 0.8615 & \textbf{0.5654} & \textbf{0.7344} & \textbf{0.5892} \\
    \midrule
    \multicolumn{5}{l}{\textit{Reference rankers}} \\
    Production ranker
      & --- & --- & --- & 0.6141 \\
    Teacher Ranker
      & --- & --- & --- & 0.6199 \\
    \bottomrule
  \end{tabular}}

  \label{tab:offline}
{\footnotesize\normalfont
Gryphon-v2 (beam order) uses the Gryphon-v2 checkpoint with the Ranking Module bypassed. Bold marks the best point estimate among the generative models. The reference rankers generate no candidates and are reported on WPA only. They are shown for context and are not matched comparisons, since they are trained on a full year of logged feedback against the two-week window used for the generative models.
}
\end{table}

\paragraph{Candidate generation.}
On R@1000, Gryphon-v2 lands between the two baselines at $0.8615$, within the metric's run-to-run standard deviation
($\approx 0.003$) of each. Gryphon-v2 with the Ranking Module bypassed shares the
same value by construction, since R@1000 is computed on the generated pool before
reranking. We read this as evidence that the distilled Ranking Module does not
come at a cost in candidate-generation quality.


\paragraph{Top-\(k\) selection fidelity.}
Without distillation, none of the orderings recover the Teacher Ranker's
selections: beam-score ordering (Generative retrieval, and Gryphon-v2 with the
Ranking Module bypassed) and Gryphon's next-item-trained Ranking Module all reach
T-R@10 below \(0.04\) and T-R@100 below \(0.20\).
Distillation changes this sharply: Gryphon-v2 reaches \(0.5654\) and \(0.7344\),
more than an order of magnitude higher at \(k{=}10\). Notably, Gryphon's Ranking
Module has the same architecture as Gryphon-v2's, so the gap is attributable to
the supervision rather than to item-level scoring per se.



\paragraph{Impression-ranking quality.}
Among the generative models, Gryphon-v2 achieves the highest WPA at
\(0.5892\), against at most \(0.5528\) for the other generative models.
We additionally report WPA for the reference rankers: the Teacher Ranker
(\(0.6199\)), which supplies the distillation targets, and the
production ranker (\(0.6141\)). Both are trained on a full year of
logged feedback, compared with the two-week window used for the
generative models, so we present them as a quality ceiling and not as
matched comparisons. Relative to that ceiling, teacher distillation in
Gryphon-v2 closes \(54\%\) of the WPA gap between next-item supervision
and the Teacher Ranker, and \(59\%\) of the gap to the production
ranker: the distilled Ranking Module recovers the majority of each gap
from a two-week training window.



\paragraph{End-to-end interpretation.}
Taken together, the offline results show that Gryphon-v2 preserves the
candidate-generation quality of the Semantic-ID backbone, raises
agreement with the Teacher Ranker on generated candidates by more than
an order of magnitude at \(k=10\), and attains the highest weighted pair
accuracy among the generative models, recovering the majority of the gap
to the reference rankers. Each metric, however, covers a different
component and a different candidate distribution: R@1000 evaluates the
generated pool before reranking, T-R@\(k\) evaluates fidelity to the
teacher on candidates the model itself generates, and WPA evaluates
ordering on impressions logged under the production cascade's exposure
policy. No single one of them measures the unified generate-and-rank
system end to end, so we treat the online A/B experiment in
Section~\ref{sec:online} as the primary assessment of user-facing
quality.

\subsection{Online A/B Experiment Results (RQ2)}
\label{sec:online}
\newcommand{\abTLT}{+1.62\%}
\newcommand{\abActiveUsers}{+1.41\%}
\newcommand{\abLikes}{+7.12\%}
\newcommand{\abSkipsRate}{-10.14\%}
\newcommand{\abDislikesRate}{+0.07\%}
\newcommand{\abRepeats}{+15.25\%}
\newcommand{\abUnfinishedRate}{9.65\%}

We validate Gryphon-v2 through an online A/B experiment in a large-scale music recommendation service.
Eligible users were assigned at the user level to disjoint control and
treatment arms, each containing \(8\%\) of the eligible population. The
control arm is the full production cascade: more than 15 heterogeneous
candidate generators produce roughly 10{,}000 candidates per request, a
production pre-ranker reduces them to 3{,}000, and a separate production ranker
orders the final slate. In the treatment arm, this entire cascade---candidate
generation, pre-ranking, and ranking---is replaced by a single Gryphon-v2 model,
configured as reported in Table~\ref{tab:hparams:backbone}: it generates a beam of \(1{,}024\) valid SIDs, expands their collision classes,
caps the resulting pool at \(1{,}200\) resolved items, and produces the final ordering with a
single-layer item-level Ranking Module that combines its heads into a ranking
score using the same combination as the Teacher Ranker --- with no separate
downstream ranker. In contrast to Gryphon, which supplied candidates to an
unchanged production ranker~\cite{tikhonovich2026gryphon}, Gryphon-v2 is the sole
model on the serving path.

Table~\ref{tab:online} reports the results; all deltas are relative to the
production cascade. The primary engagement metric is the number of active
users---users who listen for at least 7 minutes per day---which increased by
\abActiveUsers{}. Total listening time (TLT) increased by \abTLT{}, and
Gryphon-v2 also moved the secondary quality metrics favorably: the
unfinished-track ratio decreased by \abUnfinishedRate{}, likes increased by
\abLikes{}, and ``Repeat'' commands increased by \abRepeats{}. These results show that, on treatment traffic during the experiment, a single learned generate-and-rank model could replace the candidate-generation, pre-ranking, and ranking stages while significantly improving the engagement metric.

\begin{table}[t]
  \centering
  \caption{Online A/B experiment results: Gryphon-v2 versus the production cascade.}
  \label{tab:online}
  \begin{tabular}{lc}
    \toprule
    Metric & Relative change  \\
    \midrule
    Total listening time & \abTLT \\
    Active users$^\dagger$          & \abActiveUsers \\
    Likes                          & \abLikes \\
    ``Repeat'' commands            & \abRepeats \\
    Unfinished-track ratio         & -\abUnfinishedRate  \\
    \bottomrule
  \end{tabular}
  
{\footnotesize\normalfont
$^\dagger$ Primary engagement metric; active users are those with daily
listening time $\geq 7$ minutes. All reported deltas are relative to the full
production cascade and statistically significant at $p < 0.001$.
}
\end{table}

\begin{table}[t]
  \centering
  \small
  \setlength{\tabcolsep}{3pt}
  \caption{Ablations of candidate source, rollout beam size, and
  distillation loss.}
  \resizebox{\columnwidth}{!}{%
  \begin{tabular}{@{}lcccc@{}}
    \toprule
    \textbf{Configuration}
      & \textbf{R@1000}
      & \textbf{T-R@10}
      & \textbf{T-R@100}
      & \textbf{WPA} \\
    \midrule

    \multicolumn{5}{l}{\textit{Candidate source}} \\
    Rollout + impressions$^\dagger$
      & 0.8615 & \textbf{0.5654} & \textbf{0.7344}
      & \textbf{0.5892} \\
    Rollout only
      & 0.8610 & 0.5618 & 0.7288 & 0.5730 \\
    Impressions only
      & \textbf{0.8663} & 0.2983 & 0.5281 & 0.5872 \\

    \midrule
    \multicolumn{5}{l}{\textit{Training rollout beam size}} \\
    32$^\dagger$
      & 0.8615 & 0.5654 & 0.7344 & 0.5892 \\
    64
      & 0.8565 & 0.5661 & 0.7359 & 0.5901 \\
    128
      & \textbf{0.8616} & \textbf{0.5762} & \textbf{0.7427}
      & \textbf{0.5905} \\

    \midrule
    \multicolumn{5}{l}{\textit{Distillation loss}} \\
    MAE$^\dagger$
      & 0.8615 & 0.5654 & \textbf{0.7344} & 0.5892 \\
    MSE
      & 0.8671 & \textbf{0.5748} & 0.7329 & \textbf{0.5915} \\
    Huber
      & \textbf{0.8679} & 0.5568 & 0.7220 & 0.5894 \\
    KL
      & 0.8662 & 0.5452 & 0.7151 & 0.5860 \\
    \bottomrule
  \end{tabular}%
  }
  \label{tab:eval:ablations}
{\footnotesize\normalfont
Each
  block varies one factor relative to the \(^{\dagger}\)-marked online production configuration.
  Bold marks the best point estimate within each block and does
  not imply statistical significance.
}
\end{table}

\subsection{Ablations (RQ3)}
\label{sec:ablate}

We conduct post-hoc ablations of candidate source, training rollout beam size,
and distillation loss, varying each factor relative to the deployed Gryphon-v2 configuration.
Table~\ref{tab:eval:ablations} summarizes the results.

\paragraph{Candidate source.}
Rollouts draw distillation candidates from the same constrained-decoding mechanism used at serving time, though at a smaller beam width, while logged impressions broaden coverage to items with observed production exposure. Impression-only training is close to the deployed configuration on WPA but recovers only about half as much of the teacher's top-10 selections (0.2983 vs. 0.5654). Rollout-only training retains teacher fidelity but gives the lowest WPA in the table. The deployed mixture is at least as good as either single-source variant on the metric that variant favours.

\paragraph{Rollout beam size.}
The training beam provides a computationally bounded on-policy sample of the
serving pool. Matching the serving beam of \(1{,}024\) carries a prohibitive
training cost because every candidate must be generated, resolved, and scored
by the Teacher Ranker. The online experiment used beam size 32,
fixed before this post-hoc analysis. Beam sizes 64 and 128 produce modest
increases in some offline descriptive point estimates at higher compute cost.

\paragraph{Distillation loss.}
No objective dominates every metric: MSE performs best on T-R@10 and WPA,
MAE on T-R@100, and Huber on R@1000. The deployed online configuration used MAE
and was fixed before the final post-hoc ablation sweep, so the
MSE result remains an offline finding.

\subsection{Serving Efficiency and Stack Simplification}
\label{sec:efficiency}

\begin{table}[t]
  \centering
  \caption{Serving-stack complexity and latency under matched traffic: production cascade versus Gryphon-v2.}
  \label{tab:efficiency}
  \begin{tabular}{lcc}
    \toprule
    Measure & Production cascade & Gryphon-v2 \\
    \midrule
    Candidate generators   & $>15$                  & One model \\
    Pre-ranking stage      & Separate               & Integrated  \\
    Ranking stage          & Separate               & Integrated \\
    Candidates / request   & $\sim10{,}000 \to 3{,}000$ & $\leq 1{,}200$ \\
    End-to-end latency     & \multicolumn{2}{c}{Comparable} \\
    \bottomrule
  \end{tabular}
\end{table}

Beyond the online quality results, Gryphon-v2's central operational claim is
that one served model can replace the entire multi-stage cascade.
Table~\ref{tab:efficiency} compares the two serving paths under matched traffic
conditions. A generative backbone followed by the production ranker would not
provide a comparable unified alternative because the ranker's feature pipeline
depends on representations from other deep models and would retain much of the
cascade's serving cost.

\paragraph{Candidate volume.}
The production cascade generates roughly \(10{,}000\) candidates per request
and pre-ranks \(3{,}000\). Gryphon-v2 generates \(1{,}024\) valid SIDs and,
after collision expansion, ranks at most \(1{,}200\) resolved items---approximately
an order of magnitude fewer than the cascade's initial fan-out.

\paragraph{Serving cost.}
Gryphon-v2 is served on GPU using the NVIDIA Triton Inference Server
(Section~\ref{sec:deploy}). Direct request-level measurements---including
CPU-side feature construction, network transfer, inference-server overhead,
candidate generation, and ranking---place its end-to-end latency on par with the
production cascade on the same recommendation surface (Table~\ref{tab:efficiency}).
The unified path therefore matches the multi-stage cascade's serving latency while
eliminating its inter-service transitions and successive candidate-scoring stages.

\paragraph{Throughput.}
Under matched serving conditions, Gryphon-v2 sustains approximately
\(4\times\) the throughput of the same generative backbone followed online by the Teacher Ranker. It reuses its \(2{,}048\)-event history encoding in the
lightweight Ranking Module, avoiding the Teacher Ranker's independent
inference pass over up to \(8{,}000\) events.
Distillation therefore removes the Teacher Ranker's substantially more
expensive inference pass from the serving path.

\section{Limitations and Future Work}
The online A/B experiment evaluates Gryphon-v2 as an end-to-end treatment. Consequently, the observed deltas estimate the aggregate effect of replacing the production cascade and do not isolate the marginal contributions of Rollout Distillation, the Ranking Module, or online updates.

The experiment ran on a single music recommendation surface,
with disjoint user-level control and treatment arms, each containing \(8\%\) of eligible users. The results establish short-term viability on this sampled traffic but do not establish long-term effects or effects after deployment to the full eligible population. We did not evaluate long-tail catalogue coverage, artist diversity, novelty, or exposure concentration.
Measuring these effects under longer and broader deployments is important future work.

TeacherRecall@\(k\) uses the same Teacher Ranker that provides the distillation targets and therefore measures distillation fidelity. WPA evaluates ordering on logged impressions under the existing
cascade's exposure policy, and therefore does not measure ordering quality on the
items Gryphon-v2 itself surfaces. None of them measure the end-to-end independent recommendation quality.

Finally, Rollout Distillation is supervised distillation: it neither uses policy gradients nor directly optimizes the decoder against a reward. We did not conduct a controlled comparison with post-training based on reinforcement learning and therefore cannot draw conclusions about relative quality, stability, or computational efficiency. The approaches may be complementary.

\section{Conclusion}

We presented Gryphon-v2, a shared-encoder model that unifies Semantic-ID candidate generation and item-level ranking, with the Ranking Module distilled from a high-capacity, training-only Teacher Ranker. Rollout Distillation collects teacher targets over candidates generated by the current decoder together with logged impressions, improving both agreement with the Teacher Ranker and label-based ranking accuracy over generative baselines, while fully preserving retrieval recall. In an online A/B experiment, a single Gryphon-v2 model replaced the entire production cascade --- more than 15 candidate generators, pre-ranking, and final ranking --- as the only learned model on the treatment serving path, increasing the number of active users by 1.41\% at serving latency comparable to the production cascade. These results demonstrate the practical viability of a unified generate-and-rank model as an alternative to a multi-stage recommendation cascade.


\bibliographystyle{ACM-Reference-Format}
\bibliography{references}

\appendix
\section{Hyperparameters and Training Schedule}
\label{sec:hparams-training}

\paragraph{Architecture.}
Table~\ref{tab:hparams:backbone} summarizes the shared backbone configuration
and the depth of the Ranking Module.

\begin{table}[!htbp]
  \centering
  \small
  \caption{Gryphon-v2 architecture hyperparameters.}
  \begin{tabular}{lc}
    \toprule
    \textbf{Parameter} & \textbf{Value} \\
    \midrule
    Encoder layers                   & 7 (bidirectional) \\
    Decoder layers                   & 2 \\
    Ranking Module layers            & 1 \\
    Hidden dimension                 & 1{,}024 \\
    Attention heads                  & 16 \\
    Head dimension                   & 64 \\
    Max user-history length          & 2{,}048 events \\
    Number of SID codebooks          & 3 \\
    SID codebook size                & 32{,}000 \\
    \midrule
    Total parameters                 & 0.5B \\
    \bottomrule
  \end{tabular}
  \label{tab:hparams:backbone}
\end{table}

\paragraph{Offline training.}
We jointly optimize Gryphon-v2 with AdamW under FSDP2 sharding. Each GPU
processes 32 requests per microbatch; across 128 GPUs and four gradient
accumulation steps, this yields an effective batch size of \(16{,}384\)
requests. Offline training makes one chronological pass over the training
window. The learning rate increases linearly from \(10^{-5}\) to
\(3\times10^{-4}\) during the first \(3{,}000\) optimization steps and then
decreases linearly to \(7\times10^{-5}\) by the end of training.

\paragraph{Online updates.}
Serving is initialized from the offline checkpoint. Subsequent updates on fresh
interaction data use the same optimizer and effective batch size as offline
training, with the learning rate held constant at \(7\times10^{-5}\), the final
value of the offline schedule.

\section{Generative AI Usage Disclosure}
Generative AI tools were used to support manuscript revision, including English-language editing, clarity improvements, and high-level feedback on the presentation before submission. The authors manually reviewed and edited all AI-assisted suggestions and take full responsibility for the final manuscript.

\end{document}